\documentclass[aps,prb,showpacs,amsmath,amssymb,twocolumn]{revtex4-1}
\usepackage{graphicx}
\usepackage{color}

\begin{document}
\title{Mechanism of structural phase transitions in KCrF$_3$}
\author{Carmine Autieri}
\affiliation{Dipartimento di Fisica ``E.R. Caianiello'', Universit\`a di
Salerno, I-84084 Fisciano (SA), Italy}
\affiliation{Institute for Advanced Simulation, Forschungszentrum J\"{u}lich, 52425 J\"{u}lich,  Germany}
\author{Erik Koch}
\affiliation{German Research School for Simulation Science, 52425 J\"ulich, Germany}
\affiliation{JARA High-Performance Computing, RWTH Aachen University, 52062 Aachen, Germany}
\author{Eva Pavarini}
\affiliation{Institute for Advanced Simulation, Forschungszentrum J\"{u}lich, 52425 J\"{u}lich,  Germany}
\affiliation{JARA High-Performance Computing, RWTH Aachen University, 52062 Aachen, Germany}
\begin{abstract}
We study the origin of the cubic to tetragonal and
tetragonal to monoclinic structural transitions in KCrF$_{3}$,
and the associated change in orbital order, paying particular attention to the relevance of super-exchange in both phases.
We show that  super-exchange  
is not the main mechanism driving these transitions.
Specifically, it is not strong enough to be responsible for the high-temperature
cubic to tetragonal transition and does not yield the type of orbital order 
observed in the monoclinic phase. The energy difference between the tetragonal and the monoclinic
structure is tiny, and most likely results from the interplay between volume, covalency, and localization effects. 
The transition is rather driven by Slater exchange than super-exchange. Nevertheless, once the monoclinic distortions are present, super-exchange
helps in stabilizing the low symmetry structure.
The orbital order we obtain for this monoclinic phase
is consistent with the magnetic transition at 80~K.
\end{abstract}
\maketitle

\section{Introduction}
\begin{figure*}[t]
\centering
\rotatebox{270}{\includegraphics[width=0.46\textwidth]{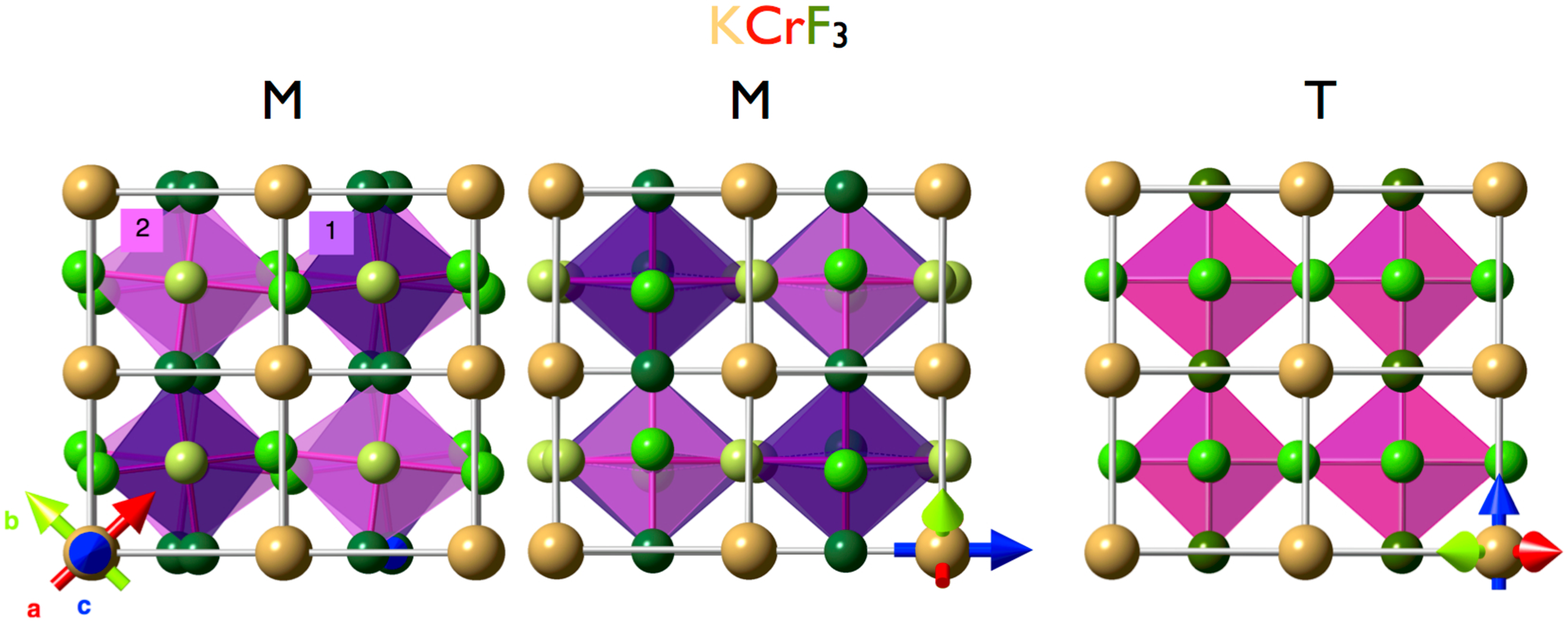}}
\caption{(Color online) \label{structure} 
The monoclinic (M, left and center) and tetragonal structure (T, right) of KCrF$_3$.
Atoms: K (large spheres), F (intermediate size spheres; different shades (light/dark) show inequivalent F atoms) forming
octahedra around Cr centers.
For the monoclinic structure the figure on the left shows the tilting of the octahedra about the ${\bf c}$ axis; the octahedra do not rotate, as shown by the central figure. 
To better illustrate these distortions, two consecutive planes in direction $\bf c$ (left figure) or ${\bf a}-{\bf b}$ (center figure) are shown.
With respect to the tetragonal structure, 
long and short Cr-F bond shrink at sites labeled as type 1 and elongate
at sites labeled as type 2. 
The pseudo-cubic directions are defined as follows.
Tetragonal structure: ${\bf x} \sim ({\bf a} + {\bf b} )/2$, ${\bf y }\sim (-{\bf a} +{\bf b})/2$, and 
${\bf z} \sim {\bf c}/2$.  
For the $i=1$ site the long (short) bond is along ${\bf x}$ (${\bf y}$) direction. 
Monoclinic structure:  ${\bf x} \sim {\bf c} /2$, 
${\bf y}\sim ({\bf a} - {\bf b})/2$ and 
${\bf z} \sim ({\bf a} + {\bf b})/2$. 
For the octahedron $i=1$ site the long (short) bond is along ${\bf x}$ (${\bf y}$) direction.
This choice of pseudo-cubic axes allows direct comparison between the structure in the two phases: The figure on the left for the monoclinic and figure on the right for the tetragonal show the same view ($yz$ plane, site of type 1 on the right top corner).}
\end{figure*}

The Mott insulator KCrF$_3$ is isoelectronic
to LaMnO$_3$, the mother compound of colossal magnetoresistance materials,
but differently from LaMnO$_3$ it exhibits a series of structural
and magnetic phase transitions.\cite{Margdonna2007,Xiao10}
At temperatures higher than 973~K it is a cubic perovskite, between 973 and 250 K it is tetragonal
and finally below 250 K it becomes monoclinic. 
The tetragonal and monoclinic structures are shown in Fig.~\ref{structure}.
At the 973~K transition, with the lowering of the symmetry from cubic to tetragonal a cooperative Jahn-Teller (JT) distortion develops.\cite{Margdonna2007} It is of    G-type
(short and long CrF bonds alternate in all directions), while in LaMnO$_3$
 the order is instead of C-type
(short and long bonds alternate in the $\bf a \bf b$ plane and repeat along the $\bf c$ direction).
Thus below 973~K the system is orbitally ordered.
Finally, KCrF$_3$ becomes
magnetic below $T_{\rm N}\sim 80$~K; the ordering vector is $(1/2\pm \delta,1/2\pm \delta,0)$, 
corresponding to an antiferromagnetic
A-type order with an incommensurate component $\delta$ which disappears at 46~K.\cite{Xiao10}
The phase transitions of KCrF$_3$ have been intensively investigated, \cite{Margdonna2007,Xiao10,Giovannetti,goodenoughnew,Xu} but
their nature, and in particular the role played by the purely electronic super-exchange mechanism
in the structural transitions, is to date not fully understood.

In recent years we have studied the origin of G- and C-type Jahn-Teller distortions
 in KCuF$_3$, 
LaMnO$_3$,
and rare-earth manganites.\cite{oo1,oo2,oo3} We have shown that, although  Kugel-Khomskii (KK) many-body super-exchange \cite{kk}
is very large, it appears to have little influence on the high-temperature orbital-order to orbital disorder transition
observed experimentally\cite{oomelting}  in the full series
of rare-earth manganites. 
However, in particular in LaMnO$_3$, super-exchange effects turned out to be so strong that, if hypothetically the
static Jahn-Teller distortion was absent, it could alone explain 
an orbital-order transition at temperatures as large as 500~K.
Remarkably, KCrF$_3$  exhibits  a change in the co-operative Jahn-Teller distortion around 250~K;
in the monoclinic phase the orbital order acquires a small C-type component
in the $yz$ plane, where the pseudocubic $\bf z$ and ${\bf y}$ axes are defined as $({\bf a}+{\bf b})/2$ and  $({\bf a}-{\bf b})/2$ (see Fig.~\ref{structure}).
Thus, super-exchange could play an important role  for the tetragonal
to monoclinic structural phase transition, or in similar low temperature
phase transitions observed in other systems.
In this work we want to clarify if that is the case.

The paper is organized as follows.
In section II we discuss the methods and models used.
In section III we present our results. 
In section III.A  we discuss the electronic
structure, obtained using density-functional theory (DFT)
in the generalized-gradient approximation\cite{notegga} (GGA). In section III.B we focus on the super-exchange mechanism
for  orbital order; 
by using the density-functional theory 
+ dynamical mean-field theory (DFT+DMFT) method,\cite{lda+dmft,book2011}
we calculate for each structure the transition temperature,
as well as the occupied orbitals.
We use the technique introduced in Ref.~\onlinecite{oo1}.
We study both the cubic to tetragonal and tetragonal to monoclinic structural phase transitions.
In section III.C we investigate the effect of the changes in volume
by using density-functional theory in the GGA
as well as the GGA+$U$ approach.\cite{Anisimov91,Anisimov93,Anisimov95}
In section III.D we discuss the origin of the magnetic structure
in the monoclinic phase.
Finally, section IV gives our conclusions.

\section{Model and method}

We calculate the electronic structure in the different phases 
and optimize the structures {\it ab-initio} using the projected augmented plane-wave
technique as implemented in the ABINIT code\cite{Abinit2008PAW,Abinit2009,Amadon2008} and in the VASP
package.\cite{VASP}  We construct Wannier functions via the Marzari-Vanderbilt
localization procedure (Wannier90 code\cite{Wannier90})
as well as via the first-principles downfolding approach based on
the $Nth$-order muffin-tin orbital (NMTO) method.\cite{NMTO}

\begin{figure}[t]
\centering
\includegraphics[width=0.52\textwidth]{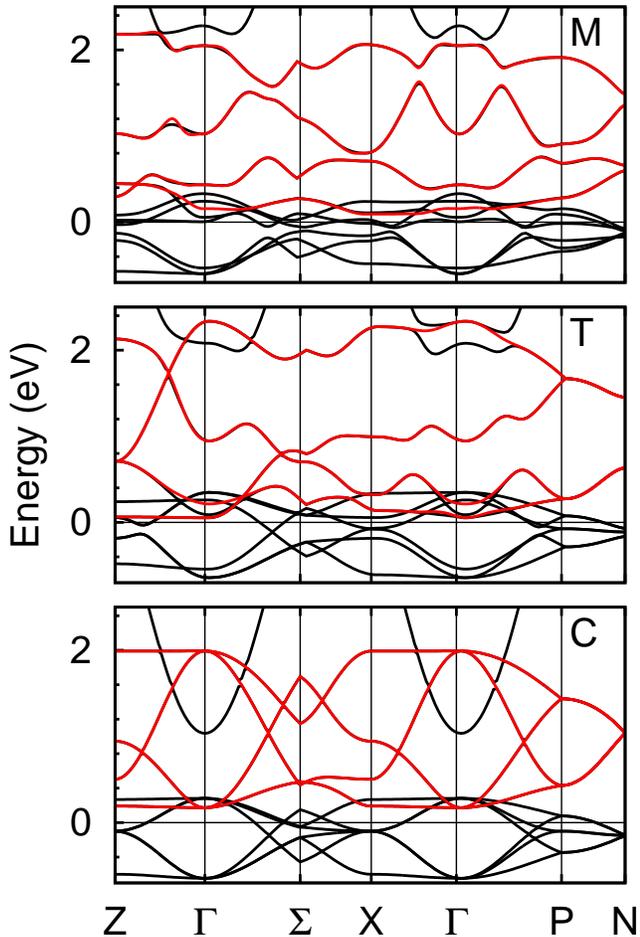}
\caption{(Color online) \label{bands} Dark lines: GGA band structure 
for the cubic (C), tetragonal (T) and monoclinic (M) phase. The Fermi level is set at the energy zero.
Light lines: $e_g$-like bands from maximally-localized Wannier functions plotted on top of the GGA bands.
The remaining bands are the Cr $t_{2g}$ bands, crossing the Fermi level and partially filled, and the empty Cr $4s$ bands.}
\end{figure}

To study the effects of the Kugel-Khomskii  super-exchange mechanism
we use {\em ab-initio} minimal many-body models.
The Cr $d$ bands split into half-filled $t_{2g}$
and 1/4-filled $e_g$ bands.
The Hund's rule interaction between $t_{2g}$ and
$e_g$ electrons yields a magnetic coupling of the $e_g$ electrons
to the effective spin of $t_{2g}$ electrons, ${\bf S}_{t_{2g}}$.
The latter acts as an effective magnetic field $h={J}S_{t_{2g}}$
and, in the paramagnetic phase, yields a band-renormalization factor accounting for 
$t_{2g}$ spin-disorder.\cite{millis}
Thus the minimal model is
\begin{eqnarray}
 \nonumber
 H &=&\!\!\!\sum_{im\sigma}\sum_{i^\prime m'\sigma'} \!\!\! 
   t^{i,i'}_{m,m'} u^{i,i'}_{\sigma,\sigma'} 
   c^{\dagger}_{im\sigma} c^{\phantom{\dagger}}_{i' m'\sigma'}\\
   \nonumber
   &-&h\sum_{im} (n_{im\Uparrow}-n_{im\Downarrow}) 
      +U\sum_{ im }  n_{im\Uparrow }n_{im\Downarrow} \\
   &+&\!\!\frac{1}{2}\!\sum_{im\left( \neq m'\right)\sigma\sigma'}
      \!\!\!(U-2J-J\delta_{\sigma,\sigma'}) n_{ im\sigma} n_{im'\sigma'}\;.
\label{H}
\end{eqnarray}
In this model $c_{im\sigma}^{\dagger}$ creates an electron with spin $\sigma\!=\Uparrow,\Downarrow$ in a Wannier orbital $|m\rangle=|x^2-y^2\rangle$ or $|3z^2-r^2\rangle$  at site $i$, and $n_{im\sigma}=c_{im\sigma}^{\dagger}c^{\phantom{\dagger}}_{im\sigma}$. $\Uparrow$ ($\Downarrow$) indicates the $e_g$ spin parallel (antiparallel) to the ${t_{2g}}$ spins on the same site. The matrix $u$ accounts for the orientational disorder  of the ${t_{2g}}$ spins,
$u^{i,i'}_{\sigma,\sigma'}=2/3$ for $i\neq i'$, $u^{i,i}_{\sigma,\sigma'}=\delta_{\sigma,\sigma'}$.
The parameter $t^{i,i'}_{m,m'}$ is the hopping integral from orbital $m$ on site $i$ to  orbital $m'$ on site $i'$. The on-site terms $t_{m,m'}=\varepsilon_{m,m'}$ give the crystal-field splitting. $U$ and $J$ are the direct and exchange  terms of the screened on-site Coulomb interaction. 
The Wannier basis provides us with {\em ab-initio} values of the hopping integrals and crystal-field splittings.
We calculate the average Coulomb interaction\cite{book2011,udefinition}
$U_{\rm av}-J_{\rm av}$ using the linear-response approach.\cite{Cococcioni05} 
We find that $U_{\rm av}-J_{\rm av}$ varies from $\sim 3$~eV in the tetragonal phases to 
$\sim 4$~eV in the monoclinic phase. The same approach yields $U_{\rm av}-J_{\rm av}\sim 2.7$~eV
for LaMnO$_3$. The theoretical estimate for $J_{\rm av}$ is $\sim$~0.75~eV.\cite{MF}
This approach leads to $U\sim U_{\rm av} +8/7J_{\rm av}\sim 5$-$6$~eV.
The GGA band structure in the different phases is shown in Fig.~\ref{bands}.

\begin{table}[t]
\begin{center}
\begin{tabular}{r|rrrrrrrrrrrrrrr}
&  \multicolumn{4}{c}{ Cubic} &  \multicolumn{4}{c}{ Tetragonal } &  \multicolumn{4}{c}{Monoclinic} \\[1ex]
lmn &  {$t^{i,i^\prime}_{1,1}$} 
    &  {$t^{i,i^\prime}_{1,2}$} 
    &  {$t^{i,i^\prime}_{2,2}$} 
    &
    &  {$t^{i,i^\prime}_{1,1}$} 
    &  {$t^{i,i^\prime}_{1,2}$} 
    &  {$t^{i,i^\prime}_{2,2}$} 
    &&  {$t^{i,i^\prime}_{1,1}$} 
    & {$t^{i,i^\prime}_{1,2}$} 
    & {$t^{i,i^\prime}_{2,1}$} 
    &  {$t^{i,i^\prime}_{2,2}$} \\
\hline
100 &-223 & 124 &  -80 && -171&  157&    -95 && -164 & 121 &   83 &   -72 \\
010 &-223 &-124 & -80  && -171& -109&   -95 && -163 & -87 & -167 &   -67 \\
001 &  -9 &   0 &  -294&&   47&  -73&    -292 &&   33 & -72 &   52 &  -253 \\[2ex]  
&  {$\varepsilon_{1,1}$} 
&  {$\varepsilon_{2,2}$} 
& {$\varepsilon_{1,2}$}
& &  {$\varepsilon_{1,1}$} 
&  {$\varepsilon_{2,2}$} 
& {$\varepsilon_{1,2}$} 
& {} 
&&  {$\varepsilon_{1,1}^{\rm Cr1}$} 
&  {$\varepsilon_{2,2}^{\rm Cr1}$} 
&  {$\varepsilon_{1,2}^{\rm Cr1}$} \\[1ex]
\hline
000 &  0 & 0 & 0 &  & 0 & 310 &  390 &   && 0 &    466 &   414 &  \\[2ex]

&&&&&&& &&
& \multicolumn{1}{c} {$\varepsilon_{1,1}^{\rm Cr2}$} 
& \multicolumn{1}{c} {$\varepsilon_{2,2}^{\rm Cr2}$} 
& \multicolumn{1}{c} {$\varepsilon_{1,2}^{\rm Cr2}$} \\[1ex]
&  &  &&  & &  &  &  & &  111  &    368 &   -316 \\[1ex]
&&&&&&& && &&& \\
\hline
$\lambda_\parallel$ & 7 &&&& 15 &&&& 34 \\
$\lambda_\perp$ & 2 &&&& 3  &&&&  4 \\
\end{tabular}
\end{center}
\caption{\label{hoppings}
Nearest neighbor hopping integrals $t^{i,i^\prime}_{m,m^\prime}$ and crystal-field matrix elements $\varepsilon_{m,m^\prime}$ in the $e_g$-like basis, with $|1\rangle=|x^2-y^2\rangle$ and
$|2\rangle=|3z^2-r^2\rangle$.
All energies are in meV. For the crystal-field levels we take $\varepsilon_{1,1}$ at site 1 as 
energy zero. The spin-orbit coupling constants $\lambda_\parallel$ and $\lambda_\perp$,
with $H_{\rm SO}=\lambda_\parallel L_zS_z +\frac{1}{2}\lambda_\perp (L_+S_-+L_-S_+)$, are also given. The directions $(lmn)$ are defined $l{\bf x}+m{\bf y} +n {\bf z}$ where
$\bf x$, $\bf y$ and $\bf z$ are the pseudo-cubic axes defined in Fig.~1. }
\end{table}

We solve the model (\ref{H}) by means of the DFT+DMFT technique.\cite{lda+dmft,book2011}
We use as impurity solver both the Hirsch-Fye \cite{hirschfye} quantum Monte Carlo (QMC) 
technique as well as the hybridization-expansion continuous-time QMC approach \cite{ctqmc}
in the implementation presented in Ref.~\onlinecite{flesch}.
We have recently shown\cite{flesch} that in $e_g$ systems such as (\ref{H})  spin-flip and pair-hopping terms
do not affect the super-exchange orbital-ordering transition temperature
$T_{\rm KK}$, and therefore we neglect them to speed up calculations.
We have also shown that the exact value of $h$ does not affect
the strength of super-exchange \cite{oo2} as far as $h$ is large enough to yield
the correct Hund's rule multiplet structure. Thus we use the theoretical estimate
for LaMnO$_3$ $h=2JS_{t_{2g}}\sim 2.7$~eV.\cite{yamasaki}

In order to study the effects of volume expansion, covalency and localization
we use the full Hamiltonian and the GGA+$U$ and SGGA+$U$ approach, where
SGGA stands for spin-polarized GGA. 
We perform calculations for different volumes, $U$ varying from 4 up to 9 eV. Finally, we calculate the magnetic coupling and the magnetic anisotropy by
combining many-body perturbation theory (based on {\em ab-initio} hopping parameters
and Coulomb integrals) and direct first principles SGGA+$U$ calculations.

\section{Results} 
\subsection{Electronic structure} 
We find that the overall bandwidth of the $e_g$ and $t_{2g}$ bands remains about the same
in all the structures, $W_{t_{2g}}\sim 1$~eV, $W_{e_g}\sim 2.3$~eV, perhaps 
$W_{t_{2g}}$ slightly decreases and $W_{e_g}$ slightly increases
reducing the symmetry. The bands themselves are, however, sizably deformed by 
the distortions, as can be seen in Fig.~\ref{bands}.

We calculate the hopping integrals and crystal-field parameters for the $e_g$ bands by constructing $e_g$ Wannier functions by projection.
The most important hopping integrals are listed in Tab.~\ref{hoppings}.
This table shows that the Jahn-Teller crystal-field splitting
progressively increases in the series of phase transitions,
while the main hopping integral, the hopping along the ${\bf z}$ direction, decreases. 
Thus, contrarily to naive expectations, the hopping integrals do not increase as the volume shrinks,
because the lattice distortions increase as well, leading to a reduction of the matrix elements due to Slater-Koster factors.
In the monoclinic case two neighboring Cr sites are inequivalent and have different splitting.
We define the lowest energy crystal-field state as
$|\theta_{\rm CF}\rangle=\cos \frac{\theta_{\rm CF}}{2} |3z^2-r^2\rangle+\sin \frac{\theta_{\rm CF}}{2}  |x^2-y^2\rangle$. Our calculations yield 
$\theta_{\rm CF}=\theta_{\rm CF}^{2}\sim  111^o=-\theta_{\rm CF}^{1}$ in the tetragonal phase, where  $\theta_{\rm CF}^{i}$ is the angle for site $i$.
In the monoclinic phase we find
$\theta_{\rm CF}^1\sim -120^o$ and
$\theta_{\rm CF}^2\sim  112^o$.
The sites and the pseudo-cubic axes are defined in Fig.~\ref{structure}.

\subsection{Kugel-Khomskii super-exchange  mechanism}
First we analyze the purely electronic Kugel-Khomskii super-exchange mechanism.
We calculate $T_{\rm KK}^{\rm T}$, the Kugel-Khomskii  critical temperature  for the transition cubic to tetragonal,
by using the approach of Ref.~\onlinecite{oo1}.
Starting from the experimental tetragonal structure we progressively reduce the Jahn-Teller and tetragonal crystal-field splitting to zero
and perform DFT+DMFT calculations for the corresponding idealized structures,
decreasing the temperature to search for the orbital order phase transition.
In the zero crystal field limit the transition is due to super-exchange only and  occurs
at a temperature $T_{\rm KK}^{\rm T}$.
We find that $T_{\rm KK}^{\rm T}\sim 400$~K, i.e., a value similar to the result
we have previously obtained for KCuF$_3$. When we define the DMFT occupied state as
$|\theta\rangle=\cos \frac{\theta}{2} |3z^2-r^2\rangle+\sin \frac{\theta}{2}  |x^2-y^2\rangle$,
we find  $\theta=-\theta_1=\theta_2\sim 90^o$, where $\theta_i$
is the angle for a site of type $i$ (see Fig.~1). The transition temperature $T_{\rm KK}^{\rm T}$ is too small for super-exchange being responsible for the high-temperature
cubic to tetragonal cooperative Jahn-Teller distortion above 900 K. 
Furthermore, the tetragonal crystal-field works against super-exchange,\cite{oo2,flesch}
leading to an occupied state with $\theta\sim 180^o$ once the 
tetragonal lattice distortions
are taken into account. This is in line with the results 
for KCuF$_3$ and REMnO$_3$ systems.\cite{oo1,oo2,oo3}
Nevertheless, $T_{\rm KK}^{\rm T}$
is sizably higher than the tetragonal to orthorhombic transition temperature, 250~K,
and thus well below $T_{\rm KK}^{\rm T}$ super-exchange could win 
and even rotate the angle defining the occupied orbital.\cite{oo2}
To verify if this is the case
we  perform a two-site cluster DFT+DMFT calculations, allowing for inequivalent
neighboring Cr sites, i.e., for the lower symmetry of the monoclinic phase. 
Surprisingly,
we find that the orbital-order transition occurs at $T_{\rm KK}^{\rm M}\sim 400~$K, i.e., 
at temperatures very similar to the critical temperature $T_{\rm KK}^{\rm T}$.
{Furthermore, we find that down to 200~K the two sites have
occupied orbitals defined by the angles $\theta =-\theta_1\sim \theta_2\sim 90^o$.
Remarkably, there is no actual big difference between the sites, suggesting that super-exchange alone cannot  account for the two inequivalent sites
in this material.
We also find a homogeneous solution in the presence of a tetragonal 
or a full (tetragonal and Jahn-Teller) crystal field. 

Finally, we perform the same cluster DMFT calculations starting directly from the monoclinic structure,
again in the absence of the static crystal field. 
Because the hopping integrals decrease  when the structure becomes monoclinic (Tab.~I), super-exchange
could become accordingly less strong; the presence of monoclinic distortions lowers, however, the
symmetry of the superexchange interaction and this could conversely by itself strongly enhance the tendency to orbital order. 
Surprisingly, we find that this is not the case. The transition temperature, $\tilde{T}_{\rm KK}^{\rm M}$, remains about the same as $T_{\rm KK}^{\rm M}$, perhaps
slightly smaller. We do find a slight site dependence of the occupied orbital, with
$\theta_1\sim -101^o$ and $\theta_2\sim 83^o$ at $\sim$ 280~K. Although apparently this goes in the correct direction,
i.e., towards the formation of inequivalent sites, the static crystal field associated with the monoclinic distortions has to be taken into account explicitly to explain the actual experimental difference
in the occupied orbital for sites of type 1 and 2. 
{The actual difference between $|\theta_1|$ and $|\theta_2|$ is explained by the larger tetragonal crystal-field splitting 
at sites of type 1 rather than by super-exchange.}
}

\subsection{Volume effect}
\begin{figure}[t]
\centering
\includegraphics[width=0.35\textwidth,angle=270]{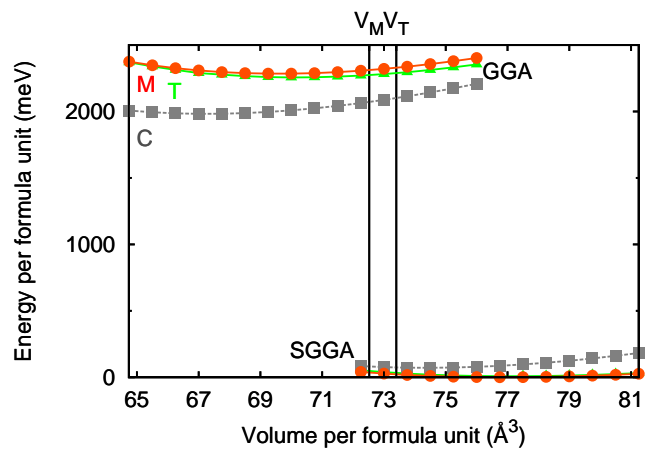}
\caption{(Color online) \label{LDA}Energy versus volume calculated in the generalized-gradient approximation (GGA)  and 
the spin-polarized generalized gradient approximation (SGGA). The  experimental
volume in the tetragonal and monoclinic case are labeled as V$_T$ and V$_M$.
 Circles: Monoclinic structure. Triangles: Tetragonal. Squares: Cubic.
 The lowest energy point is taken as the energy zero.}
\end{figure}
\begin{figure}[t]
\centering
\includegraphics[width=0.37\textwidth,angle=270]{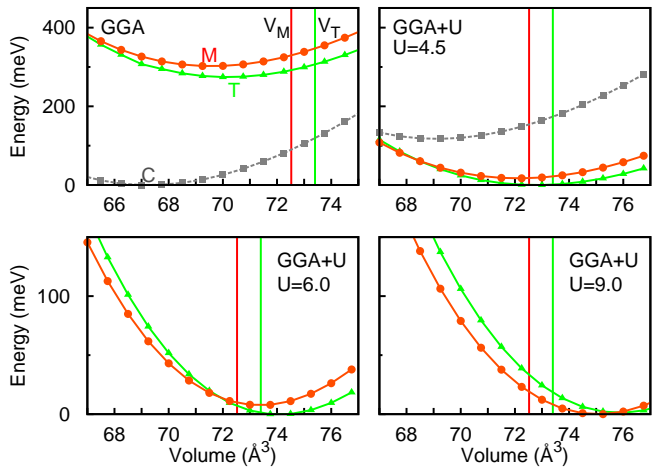}%
\caption{\label{lda+u} 
 Energy per formula unit versus volume from  GGA+$U$ for increasing $U$. The energy zero corresponds in each case
 to the lowest energy point. The labels V$_T$ and V$_M$ indicate the experimental volume 
 in the tetragonal and monoclinic structure.
 Circles: Monoclinic structure. Triangles: Tetragonal. Squares: Cubic.
 At $U\sim9$~eV the monoclinic structure becomes the lowest in energy.
 The volume is changed by uniformly scaling the unit cells.}
\end{figure}
A very different mechanism to which  tilting and rotations in perovskites
can be ascribed is the volume reduction with decreasing temperature; perhaps the tetragonal to monoclinic
transition and the associated changes in the co-operative
Jahn-Teller distortion can be explained by this phenomenon alone,
without invoking strong correlation effects. Cation covalency can further help the stabilization
of lower symmetry structures.\cite{NMTO}
To clarify whether the 250~K transition is volume- and covalency-driven we compare
the total energy of the different structures as a function of the volume. 
In Fig.~\ref{LDA} we show the total energy curves obtained
in GGA and SGGA. The GGA solutions are metallic. Having the largest hopping matrix elements of the three structures, the cubic structure is lowest in energy. The equilibrium volume is quite small as bringing the atoms closer together increases the hopping. Allowing for spin-polarization the situation changes drastically. Exchange effects open a gap and lower all energy curves by about $\sim 2$~eV. More importantly, in SGGA the cubic structure is now energetically above the other structures. In the absence of a crystal-field splitting the orbital polarization, and hence the gain in exchange energy, is smaller than in the low-symmetry phases. 
To confirm this effect, we study the different structures in GGA+$U$, changing the volume by uniformly scaling the unit cell. As shown in Fig.~\ref{lda+u}, with increasing $U$ the cubic structure becomes less and less favorable, as the orbital polarization of the insulating solution increases. 
We also observe that the position of the minimum in the energy curves shifts with increasing $U$ to larger volumes. 
The reason is that for larger $U$ the $d$-electrons tend to spread out to reduce their Hubbard energy. Thus the effective radius of the Cr ion, and therefore also the Cr--F equilibrium distance, increases with $U$. On the other hand, the effective K and F ionic radii, not involving any $d$ electrons, do not change much. Consequently, with increasing $U$ the tolerance factor decreases, favoring the tilting of the octahedra. I.e., with increasing $U$ the monoclinic structure becomes more and more favorable.  
Overall, for a given volume, the tetragonal and
monoclinic structure are very close in energy; in GGA the difference in energy
$\Delta E_{V}=E_{\rm M}(V)-E_{\rm T}(V)$ is positive and $\sim 30$-$40$~meV for volumes $V$ in the region between the GGA minima and the experimental
volumes; $\Delta E_{V}$ becomes even smaller in GGA+$U$.

Let us compare this to the super-exchange energy-gain associated with orbital order, $-\Delta E_{\rm KK}\sim k_BT_{\rm KK}/2\sim 20$~meV, with the energy differences between the various structures shown in Fig.~\ref{lda+u} calculated in GGA. 
First we consider the energy difference between the tetragonal/monoclinic structures  on the one hand and the cubic structure on the other;   $|\Delta E_{\rm KK}|$ is an order of magnitude smaller than this energy difference, which is about (in absolute value) 200-300~meV. Thus  $|\Delta E_{\rm KK}|$ alone cannot stabilize the tetragonal/monoclinic with respect to the cubic structure.
This energy gain is rather associated with the static crystal field splitting, which is $\sim 840$~meV in the tetragonal case, and the associated gain in exchange energy from orbital polarization.
Next, we consider the GGA energy difference between the monoclinic and the tetragonal structure,  $\Delta E_{V}$.
We have to compare it with the difference in orbital-order energy gain of the monoclinic structure with respect to the tetragonal structure, $\delta\Delta E_{\rm KK}$.  
Our results show that $|\delta\Delta E_{\rm KK}|$ is sizably smaller than $|\Delta E_V|$; it even has the wrong sign, i.e., $\delta\Delta E_{\rm KK}$ is positive rather than negative because $\tilde{T}_{\rm KK}^{\rm M}$ is slightly smaller than $T_{\rm KK}^{\rm T}$, and therefore would rather stabilize the tetragonal than the monoclinic structure.
Thus  Fig.~\ref{lda+u} makes clear that it is rather the {degree of localization} and the corresponding change in the equilibrium Cr--F distance which controls the relative stability of the monoclinic and tetragonal structures.

\begin{figure}[t]
\centering
\includegraphics[width=0.20\textwidth,angle=270]{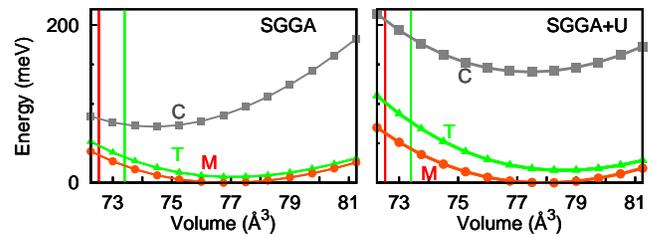}
\caption{\label{lsdau}Energy per  formula unit versus volume  from SGGA+$U$ calculations.
The ground state is shifted at zero energy.
The SGGA+$U$ calculations are for $U=6$~eV.
Circles: Monoclinic structure. Triangles: Tetragonal. 
Squares: Cubic. The vertical lines indicate the experimental volumes.
For each structure all structural parameters are optimized.
}
\end{figure}
If we also allow for spin-polarization, we obtain the SGGA+$U$ results shown in Fig.~\ref{lsdau}.
Other than in the preceding calculations we do no longer rescale the unit cell, but optimize all cell parameters that, given the space group, can be varied. Consequently, we now find that the structure with the higher symmetry is always above the structure with a lower symmetry.
All spin-polarized  calculations yield an insulating ground state for all considered volumes.
For the same reasons as discussed above, with increasing $U$ the relative energy of the cubic structure increases as does the volume at which the total energy curves have their minimum.
Since tilting the octahedra reduces the energy for small volumes, the monoclinic structure has its minimum at smaller volumes than the tetragonal. The energy (and structural) difference between the two become negligible for increasing volumes. This is in line with the observed structural transition.
SGGA without $U$ fails to reproduce the experimental ${c}/{a}$ ratio in the monoclinic phase, but the agreement is recovered in SGGA+$U$ calculations with {realistic $U\sim 5$-$6$~eV.}
Remarkably, the energy gain from lowering the symmetry from tetragonal to monoclinic, $\Delta E_{V}$, is tiny, $\sim -10$~meV in SGGA and $\sim-15$~meV in SGGA+$U$ with $U\sim 6$~eV. This is in line with a tetragonal to monoclinic transition at temperatures as low as 250~K.

\begin{table}[t]
\begin{center}
\begin{tabular}{r|rrrrrrrrrrrrrrr}
&  \multicolumn{4}{c}{ Cubic} &  \multicolumn{4}{c}{ Tetragonal } &&  \multicolumn{3}{c}{Monoclinic} \\[1ex]
lmn &  {$t^{i,i^\prime}_{1,1}$} 
    &  {$t^{i,i^\prime}_{2,2}$} 
    &  {$t^{i,i^\prime}_{3,3}$} 
    &
    &  {$t^{i,i^\prime}_{1,1}$} 
    &  {$t^{i,i^\prime}_{2,2}$} 
    &  {$t^{i,i^\prime}_{3,3}$} 
    &&&  {$t^{i,i^\prime}_{1,1}$} 
    & {$t^{i,i^\prime}_{2,2}$} 
    & {$t^{i,i^\prime}_{3,3}$} 
       \\
\hline
100 &-102 & -3   &  -102 && -94&  -3&    -89    &&& -113 & 0 &   -132 \\
010 &-102 &-102  &  -3   && -94& -89&   -3      &&&  -75 & -88 &  0     \\
001 &  -3 & -102 &  -102     &&  -6&  -144&    -144 &&&   -1 & -91 &   -86    \\[2ex]  
&  {$\varepsilon_{1,1}$} 
&  {$\varepsilon_{2,2}$} 
& {$\varepsilon_{3,3}$}
& &  {$\varepsilon_{1,1}$} 
&  {$\varepsilon_{2,2}$} 
& {$\varepsilon_{3,3}$} 
& {} 
&&  {$\varepsilon_{1,1}^{\rm Cr1}$} 
&  {$\varepsilon_{2,2}^{\rm Cr1}$} 
&  {$\varepsilon_{3,3}^{\rm Cr1}$} \\[1ex]
\hline
000 &  0 & 0 & 0 &  & 0 & 43 &  -86 &   && 0 &    -70 &   -96 &  \\[2ex]

&&&&&&& &&
& \multicolumn{1}{c} {$\varepsilon_{1,1}^{\rm Cr2}$} 
& \multicolumn{1}{c} {$\varepsilon_{2,2}^{\rm Cr2}$} 
& \multicolumn{1}{c} {$\varepsilon_{3,3}^{\rm Cr2}$} \\[1ex]
&  &  &&  & &  &  &  & &  -123 &    -29 &   -3 
\end{tabular}
\end{center}
\caption{\label{hoppingst2g}
Largest nearest neighbor hopping integrals $t^{i,i^\prime}_{m,m^\prime}$ and crystal-field matrix elements $\varepsilon_{m,m^\prime}$ in the $t_{2g}$-like basis, with $|1\rangle=|xy\rangle$,
$|2\rangle=|yz\rangle$ and $|3\rangle=|xz\rangle$.
All energies are in meV. For the crystal-field levels we take $\varepsilon_{1,1}$ at site 1 as 
energy zero. The directions ${\bf x}=(100)$, ${\bf y}=(010)$ and ${\bf z}=(001)$ are defined in the caption of Fig.~1.
 } 
\end{table}

As we have seen, orbital many-body super-exchange appears to affect hardly this energy balance.
Even a difference in energy as small as $10$~meV would correspond to a temperature difference
$T_{\rm KK}^{\rm T}-\tilde{T}_{\rm KK}^{\rm M}\sim 2\delta\Delta_{\rm KK}/k_B$ of the order of 200~K, whereas our results indicate
that the super-exchange transition temperature is about the same in the monoclinic and tetragonal phase,
and has furthermore the incorrect sign ($\delta\Delta_{\rm KK}>0$).
The difference $|\delta\Delta_{\rm KK}|$ could increase if
the screened Coulomb repulsion integral $U$ would be very different for the monoclinic and tetragonal structure.
Even if the Coulomb repulsion is slightly different in the two phases, however,
it is unlikely that it is reduced by $ 50\%$ in the monoclinic phase, as would be required to
explain a monoclinic ground state within super-exchange. Furthermore our {\em ab-initio} estimates of
$U$ indicates that this parameter is slightly larger in the monoclinic than in the tetragonal phase;
such a difference would lead again to a positive rather than negative  $\delta\Delta_{\rm KK}$,
reinforcing the conclusion that super-exchange alone does not explain the tetragonal to
monoclinic transition.

On the other hand, in the presence of static distortions a redistribution of orbital occupations follows, and it is strongly enhanced by the Coulomb repulsion; this can further stabilize the low-symmetry structures with respect
the cubic one. The $e_g$ crystal-field splitting
is modified from $\sim 840$~meV in the tetragonal structure to $\sim 950$~meV (site 1)
and $\sim 680$~meV (site 2). 
{Our DMFT calculations show that, in the presence of such crystal-field splittings,
the occupation at a temperature as high as 550~K is already basically complete
for both the tetragonal and for the monoclinic structure, differently than in GGA. Our cluster DMFT results indicate that there is no
sizable charge disproportionation, despite the difference in crystal-field
splitting between sites of type 1 and 2.}

\subsection{Magnetic superexchange}

In this last section we analyze the magnetic structure. The aim
is to verify if the change in orbital order resulting from our calculations
for the experimental structure
can explain the observed magnetic order in the monoclinic phase.
To do this we calculate the magnetic coupling
using super-exchange theory  in the basis of the Wannier functions. \cite{oo1,lisivo}
The magnetic coupling has contributions from both the half-filled
$t_{2g}$ shell and the $e_g$ shell; in a basis of orthogonal
Wannier functions we can split the two contributions so that
$J^{i,i^\prime}= J^{i,i^\prime}_{e_g}+J^{i,i^\prime}_{t_{2g}}$.
Then, if we neglect spin-flip and pair-hopping terms we arrive
at the approximate expressions
\begin{figure}[t]
\rotatebox{270}{\includegraphics[width=0.35\textwidth,angle=0]{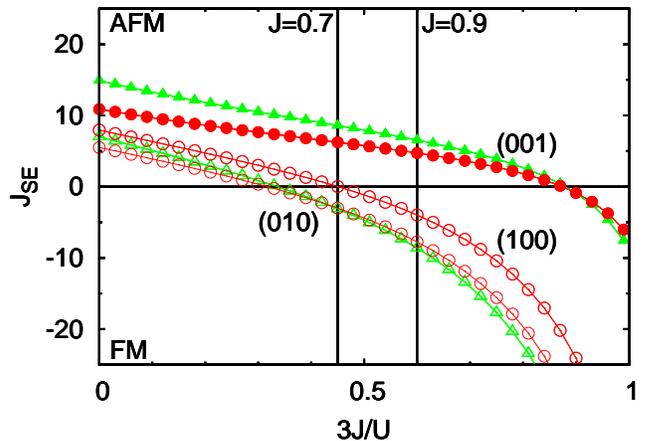}}
\caption{
\label{JSE}Super-exchange parameters as function of $3J/U$ for $U=6$~eV.
Triangles: Tetragonal structure. Circles: Monoclinic
structure. Full symbols: coupling along the ${\bf z}$ axis.
Empty symbols: coupling along ${\bf x}$ and ${\bf y}$.
The directions $\bf x$, $\bf y$ and $\bf z$ are defined in the caption of Table I.
The two vertical lines indicate realistic values of the $3J/U$ ratio.
In this range super-exchange yields an A-type antiferromagnetic structure,
in agreement with experiments.}
\end{figure}
%
\begin{eqnarray*} \nonumber
  J^{i,i^\prime}_{e_{g}}&\sim& \frac{|t_{a,a}^{i,i^\prime}|^2}{U+3J+\varepsilon_{a}^i-\varepsilon_{a}^{i^\prime}}
   + \frac{|t_{a,a}^{i,i^\prime}|^2}{U+3J+\varepsilon_{a}^{i^\prime}-\varepsilon_{a}^{i}} \\ \nonumber
                 &-&\frac{|t_{a,b}^{i,i^\prime}|^2}
                    {U-3J+\varepsilon_{b}^{i^\prime}-\varepsilon_{a}^{i}}
            \frac{4J}{U+J+\varepsilon_{b}^{i^\prime}-\varepsilon_{a}^{i}} 
              \\ \nonumber
                 &-&\frac{|t_{a,b}^{i^\prime,i}|^2}
                     {U-3J+\varepsilon_{b}^{i}-\varepsilon_{a}^{i^\prime}} 
            \frac{4J}{U+J+\varepsilon_{b}^{i}-\varepsilon_{a}^{i^\prime}} 
            ,\\[2ex]\nonumber       
  J^{i,i^\prime}_{t_{2g}}&\sim&2\frac{|t_{c,c}^{i,i^\prime}|^2 + |t_{d,d}^{i,i^\prime}|^2+|t_{e,e}^{i,i^\prime}|^2}{U+3J}.
\end{eqnarray*}
Here we denote with $|a\rangle$ and $|b\rangle$ the $e_g$ crystal-field states and
with $|c\rangle$, $|d\rangle$, $|e\rangle$ the $t_{2g}$ crystal-field states;
We find that $|c\rangle \sim |xy\rangle$, $|d\rangle \sim |yz\rangle$, $|e\rangle \sim |xz\rangle$.
{Since for the $t_{2g}$ states we find that the inter-orbital hopping integrals
are very small, for simplicity we set them to zero  in the formula above;
for the same reason we set to zero the energy difference between crystal-field
orbitals at different sites, which is at most 120 meV and leads to small corrections of order $(t^2/U)(\Delta\varepsilon/U)^2$.
The calculated exchange couplings (including also the small contributions neglected in
the analytic expression above) are shown in Fig.~\ref{JSE}. 
This figure shows that if the tetragonal structure would persist at low temperature,
the magnetic structure would be ferromagnetic and isotropic
in the $xy$ plane, and antiferromagnetic along the ${\bf z}$ axis.
In the monoclinic structure the coupling in the $xy$ plane 
remains ferromagnetic, with the ferromagnetic coupling
slightly anisotropic, because the inter-orbital $t_{2g}$ hoppings are
small, hence the antiferromagnetic contribution dominates.
Remarkably, ferromagnetism in the $xy$ plane can then be ascribed to orbital-order in
the $e_g$ states alone. On the other hand the $t_{2g}$ states
are essential for the antiferromagnetic order along $\bf z$.
All this is in excellent agreement with experiment.
Thus the orbital-order obtained in our calculation supports the
experimentally reported magnetic structure.
{
Finally, by comparing crystal-field energies with and without spin-orbit interaction, we obtain the spin-orbit couplings  (Tab.~I) and find them small 
in all systems, but larger in the monoclinic  than in the tetragonal or cubic structure.
Thus we additionally perform SGGA+$U$ magneto-crystalline anisotropy calculations and find 
that a spin orientation in the ${\bf xy}$ plane is favored, in line with experiments;\cite{Xiao10}
our  results suggest ${\bf y}$ as easy axis,  but the energy difference between ${\bf y}$ and  ${\bf x}$
is tiny (0.03 meV).}

\section{Conclusion}
We have studied the origin of orbital-order and structural phase transitions in KCrF$_3$,
a system which is isoelectronic to LaMnO$_3$. 
We could reproduce the experimental orbital- and spin-order in all phases.
We show that the Kugel-Khomskii super-exchange mechanism is not strong enough
to drive the high-temperature cubic to tetragonal transition reported at 973~K.
The tetragonal to monoclinic transition is more tricky, because
the super-exchange transition temperature is larger than the
structural transition temperature. By using the cluster DFT+DMFT approach we show, however,
that super-exchange does not support the
experimental type of orbital order in the monoclinic phase.
Next we analyze the stability of the various phases as a function
of volume. We show, by using GGA+$U$ and SGGA+$U$, that the tetragonal phase
is favored at larger volumes and the monoclinic
at small volumes, in agreement with experiments.
The difference in energy is small, $\sim 10$-$20$~meV, again in agreement
with experiments. The exact volume of the transition from tetragonal to
monoclinic depends on $U$ and the spin polarization.
Increasing $U$ the transition happens at larger and larger volumes.
The change in structure
is thus helped more by Slater exchange than by super-exchange; a triggering factor could
be a slightly larger direct Coulomb repulsion integral $U$ in the monoclinic structure.
Once the distortions are in place, DMFT calculations show that the
orbital polarization is enhanced by Coulomb repulsion, likely providing a positive feedback to
the stabilization of the distorted structure.

\acknowledgments
Calculations have been done on the J\"ulich Blue Gene/Q and Juropa. 
C.A.\ acknowledges financial support from the Fondazione Angelo Della Riccia and discussions with Mario Cuoco.
E.P.\ and E.K.\ acknowledge financial support from the Deutsche Forschungsgemeinschaft through research unit FOR 1346.


\begin{thebibliography}{99}
\bibitem{notegga} Using the local-density approximation (LDA) instead of the GGA does not alter our conclusions.
\end{thebibliography}
\end{document}